\documentclass[prd,amsfonts,amsmath,twocolumn,nofootinbib,superscriptaddress]{revtex4}

\usepackage{nicefrac}

\usepackage{color}
\usepackage{graphicx} 
\usepackage{slashed}
\usepackage{epstopdf}%

\newcommand {\be} {\begin{equation}} 
\newcommand {\ba}{\begin{eqnarray}} 
\newcommand {\ee} {\end{equation}} 
\newcommand{\ea} {\end{eqnarray}}

\renewcommand{\epsilon}{\varepsilon}

\begin{document}

\title{Constraints to new physics models for the proton charge radius puzzle from the decay $K^+ \rightarrow \mu^+ +\nu + e^- + e^+$ }

\author{Carl E. Carlson}

\affiliation{Department of Physics, College of William and Mary, Williamsburg, VA 23187, USA}

\author{Benjamin C. Rislow}

\affiliation{Department of Physics, College of William and Mary, Williamsburg, VA 23187, USA}

\affiliation{PRISMA Cluster of Excellence, Institut f\"ur Physik, Johannes-Gutenberg-Universit\"at, D-55099 Mainz, Germany}

\date{October 9, 2013}

\begin{abstract}
A possible explanation for the discrepancy between electronic and muonic hydrogen measurements of the proton charge radius are new, lepton-universality violating interactions.  Several new couplings and particles have been suggested that account for this discrepancy.  At present, these explanations are poorly constrained. Experiments such as the upcoming kaon decay experiment at JPARC may constrain or eliminate some explanations by sensitivity to the decay channel $K^+ \rightarrow \mu^+ +\nu + e^- + e^+$.  We calculate the predicted contributions of the various explanations to this channel.  The predicted signals, if present, should be large enough to be resolved in the experiment.

\end{abstract}

\maketitle



\section{Introduction}			\label{sec:intro}


The recent extraction of the charge radius from muonic hydrogen~\cite{Pohl:2010zza,Antognini:1900ns} is $7\sigma$ smaller than the electronically-determined CODATA value~\cite{Mohr:2012tt}.  This discrepancy has inspired investigations into new physics explanations.  To bring the extracted charge radius from muonic measurements into agreement with the electronic determination, any new physics explanation must lower the muonic Lamb Shift by 310 $\mu$eV.

Jaeckel and Roy~\cite{Jaeckel:2010xx} showed that the popular $U(1)'$ dark photon model cannot account for the discrepancy since it would have a larger effect on the Lamb Shift of electronic hydrogen.  An explanation involving new particles must include larger couplings to muons than electrons (the  interactions must violate lepton-universality).  Explanations must also respect the constraint placed by discrepancy between the measured and calculated value of the muon's anomalous magnetic moment, $(g-2)_\mu$.  Constructing a model that simulatanously accounts for both discrepancies is challenging because the fractional $(g-2)_\mu$ discrepancy is orders of magnitude smaller than the proton charge radius discrepancy.

Several lepton-universality explanations have been proposed.  Tucker-Smith and Yavin~\cite{TuckerSmith:2010ra} suggested the existence of a low mass particle with either scalar or vector couplings to muons and protons.  The coupling strength was set to explain the radius puzzle and the small mass was needed to respect the constraint placed by the $(g-2)_\mu$ discrepancy.  Barger~\textit{et al.}~\cite{Barger:2010aj,Barger:2011mt} showed that explanations involving scalar, vector, and tensor couplings face stringent constraints placed by meson decays.  Batell~\textit{et al.}~\cite{Batell:2011qq} modified the dark photon model by inserting an additional coupling to right-handed muons.  Though this explanation receives contributions from a well-motivated, gauge-invariant model, it also requires the existence of scalar particles to respect the $(g-2)_\mu$ constraint.  The right-handed coupling to muons also spoils the gauge-invariance.

Our explanations~\cite{Carlson:2012pc} for the proton charge radius introduced new particles with scalar and pseudoscalar (polar and axial vector) couplings to muons and protons.  The strength of the scalar (polar vector) coupling was chosen to ``explain" the proton charge radius puzzle.  The pseudoscalar (axial vector) coupling enters the expression for the anomalous magnetic moment with an opposite sign of the scalar (polar vector) coupling and was fine-tuned to ``explain" the $(g-2)_\mu$ discrepancy.  We also use the upper limit to $K\rightarrow \mu + invisible$ measured by Pang~\textit{et al.}~\cite{Pang:1989ut} (see also~\cite{Barger:2011mt}) to constrain the masses of these new particles.  This experimental limit does not constrain the Batell~\textit{et al.}~explanation since their modified dark photon decays quickly enough into an $e^+e^-$ pair.

All of these charge radius puzzle explanations should be viewed as proof-of-concepts rather than completed models.  The explanations must be confronted with more experimental constraints before earning serious consideration as realistic models.  Tests of the models will include seeking the new interactions via new radiative decays to processes involving muons.   The upcoming E36 experiment at the Japan Proton Accelerator Research Complex (JPARC) will measure kaon decays in muonic channels~\cite{Djalali:2012zz}.  In this work, after some preliminary remarks, we calculate the predicted signal of the Batell~\textit{et al.}~explanation in the decay channel $K^+ \rightarrow \mu^+ +\nu + e^- + e^+$.  We also calculate the prediction of our model when we modify it to include small couplings to electrons.  The JPARC experiment should be sensitive to these signals and its results will constrain or eliminate these explanations.

\vskip -196 mm
\hfill MITP/13-062
\vskip 199 mm


\section{QED contribution to $K^+ \rightarrow \mu^+ +\nu + e^- + e^+$.}			\label{sec:JPARC_QED}


The main goal of the E36 experiment is to measure the ratio $\Gamma(K^+\rightarrow \mu^+ +\nu_\mu)/\Gamma(K^+\rightarrow e^+ +\nu_e)$.  They expect to see a large number of kaon decays, $N(K^+\rightarrow \mu^+ +\nu_\mu) = 10^{10}$ events~\cite{Jparc}.  The branching ratio for this decay channel is 0.6355~\cite{Beringer:1900zz}.  

\begin{figure*}[htbp]
\begin{center}

\includegraphics{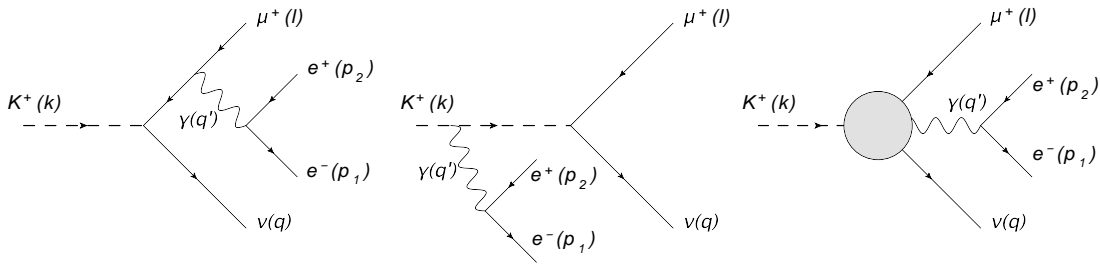}

\caption{QED contribution to $K^+\rightarrow \mu^+ +\nu_\mu +e^+ +e^-$.}
\label{kmunuee}
\end{center}
\end{figure*}

New physics explanations for the proton radius puzzle can be tested in the decay channel $K^+\rightarrow \mu^+ +\nu_\mu+e^++e^-$.  Fig.~\ref{kmunuee} shows the lowest order diagrams for standard model QED contributions to this process.  The full standard model branching ratio for this decay channel is calculated to be $2.49\times 10^{-5}$~\cite{Bijnens:1992en}.  (Independent calculations by us and by the authors of~\cite{Beranek:2013yqa} agree with this number; for an early calculation, see~\cite{Krishna:1972za}).  For their expected 50\% acceptance of $e^+e^-$ coincidences, the total number of such decays E36 can expect is
\begin{align}
N&(K^+\rightarrow \mu^+ +\nu_\mu+e^++e^-)
\nonumber \\
&=\frac{1}{2}\frac{\Gamma(K^+\rightarrow \mu^+ +\nu_\mu+e^++e^-)}{\Gamma(K^+\rightarrow \mu^+ +\nu_\mu)}N(K^+\rightarrow \mu^+ +\nu_\mu)
\nonumber \\
&\approx 2\times 10^5.
\end{align}


If no new particles are present, there will be about 1000 such events per bin, for a bin width of 1 MeV, in the vicinity of $m_{ee}=30$ MeV.  Here,   $m_{ee}$ is the energy of the $e^+e^-$ pair.  Choosing the normalization $
\langle 0 | \ \bar u  \, \gamma_\mu (1-\gamma_5) s \ | K \rangle 
	=  \sqrt{2} f_K k_\mu$, the amplitude for the QED prediction is
\begin{align}
\label{keedecay}
i{\cal M}&=-{G_F}(-ie)^2V_{us}\bar{u}(p_1)\gamma_\rho v(p_2)\frac{-i}{q'^2}(f_K m_\mu L^\rho-H^{\rho \nu}j_\nu)
\end{align}
where
\begin{align}
\label{lrho}
L^\rho &=\bar{u}(q)(1+\gamma^5)\bigg(\frac{(2k-q')^\rho}{2k \cdot q'+q'^2}-\frac{2l^\rho+\!\not\!{q'}\gamma^\rho}{2l \cdot q'+q'^2}\bigg)v(l),
\nonumber \\
H^{\rho \nu}&=-iV_1\epsilon^{\rho \nu \alpha \beta}q'_\alpha k_\beta-A_1(q'\cdot(k-q')g^{\rho \nu}
\nonumber \\
&-(k-q')^\rho q'^\nu)-A_2(q'^2 g^{\rho \nu}-q'^\rho q'^\nu),
\end{align}
and
\begin{align}
j_\nu &=\bar{u}(q)\gamma_\nu (1+\gamma^5)v(l).
\end{align}

If the kaon were pointlike, $H^{\rho\nu}$ would be zero.  In the energy regions examined in this work, the Inner Brehmsstrahlung (IB) term, $L^\rho$, dominates while $H^{\rho\nu}$ contributes less than $1\%$ to the plots.   Actual values for $V_1$, $A_1$, $A_2$, and $f_K$ can be found in Poblaguev \textit{et al.}~\cite{Poblaguev:2002ug}.  $V_1$, $A_1$, $A_2$ are given by
\begin{align}
-\sqrt{2}m_K (A_1,A_2,V_1)&=(F_A,R,F_V),
\end{align}
where $F_A=0.031$, $R=0.235$, and $F_V=0.124$.  Their value for the decay constant maps to $f_K=113$ MeV.


\section{Dark Photon contribution to $K^+ \rightarrow \mu^+ +\nu + e^- + e^+$.}			\label{sec:JPARC_Dark}


Before discussing the predictions of the Batell~\textit{et al.} model, we consider the original dark photon model it modifies.  A simple extension to the Standard Model, dark photons are the quanta of a hypothetical $U(1)'$ gauge field that interact with fermions through kinetic mixing with the $U(1)$ field associated with weak hypercharge.  
Determining constraints on the dark photon mass and its effective coupling to fermions is an active area of research and it is worth exploring whether JPARC can help this cause.

As the name suggests, the dark photon shares many of the properties of the QED photon.  If it exists, the dark photon would contribute to the decay $K^+ \rightarrow \mu^+ +\nu + e^- + e^+$. 
The dark photon's contribution to this decay is found by modifying the photon's propagator and coupling:
\begin{align}
\frac{-i}{q'^2}  &\rightarrow \frac{-i}{q'^2-m_{A'}^2+im_{A'}\Gamma},
\\
-ie &\rightarrow  -i\epsilon e.
\end{align}
In the above expressions $m_{A'}$ is the mass of the dark photon, $\Gamma$ is its decay rate into  $e^+ +e^-$ (assumed to be the dominant decay), and $\epsilon e$ is its effective coupling to fermions.  If a dark photon exists, a bump in the data will be centered around the propagator's pole and will determine the dark photon's mass.  The size of the deviation will indicate the value of $\epsilon$.  Of course, a lack of observed deviation from the Standard Model will place constraints on $m_{A'}$ and $\epsilon e$.

The square of the propagator produces a sharp peak.   When the experimental resolution is larger than the natural width of the state, as we expect here, one will observe a peak that is smeared into a Gaussian in ${q'}^2 = m_{ee}^2$, with a width given by the experimental resolution.   To simulate the experimental spread in our decay rate calculation, we replaced the Lorentzian with a Gaussian.  The normalization for the Gaussian was chosen to produce the same total decay rate when integrated over ${q'}^2$:
\begin{align}
\int d{q'}^2 &\frac{1}{({q'}^2-m_{A'}^2)^2+m_{A'}^2\Gamma^2}
\nonumber \\
&= \frac{\pi}{m_{A'}\Gamma}\int d{q'}^2 \frac{1}{2q'}\frac{1}{\sigma \sqrt{2\pi}}e^{\frac{(q'-m_{A'})^2}{2\sigma^2}},
\end{align}
where $\sigma = 1$ MeV is the bin width.

\begin{figure}[htbp]
\begin{center}

\includegraphics[width = 80mm]{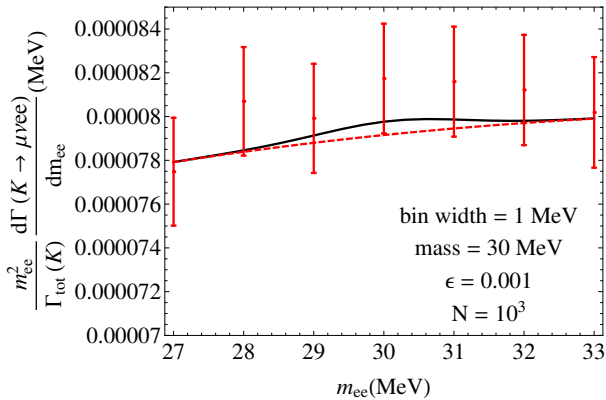}

\caption{QED prediction for the normalized differential decay rate of $K^+\rightarrow \mu^+ +\nu_\mu +e^+ +e^-$ (red, dashed curve) and the prediction with an additional dark photon (black curve).  Data points are simulated and possess fluctuations and error bars accounting for the statistical uncertainty given the number $N=1000$ anticipated events per bin.}
\label{darkphoton}
\end{center}
\end{figure}

As an example using JPARC's rate and an experimental resolution of 1 MeV, we calculated the signal of a dark photon with the currently allowed~\cite{Beranek:2013yqa} parameters $m_{A'}= 30$ MeV and $\epsilon=10^{-3}$.  The overall branching ratio for $K\to \mu \nu A' \to \mu \nu e^+ e^-$ is of order $10^{-9}$ for these parameters.  The result is shown in Fig.~\ref{darkphoton}.  The dotted red curve is the expected signal from QED while the black curve is the signal due to QED and an additional dark photon.  The simulated data points possess error bars accounting for the statistical uncertainty of 1000 events per bin.  Given the relative size of the bump and error bars, it will be very hard for JPARC to detect the presence of low mass dark photons, at least with present statistics~\cite{Beranek:2013yqa}.   However, we continue to more compelling possibilities.


\section{Batell~\textit{et al.} modified Dark Photon contribution to $K^+ \rightarrow \mu^+ +\nu + e^- + e^+$.}			\label{sec:JPARC_Batell}

\begin{figure*}[htbp]
\begin{center}

\includegraphics[width = 70mm]{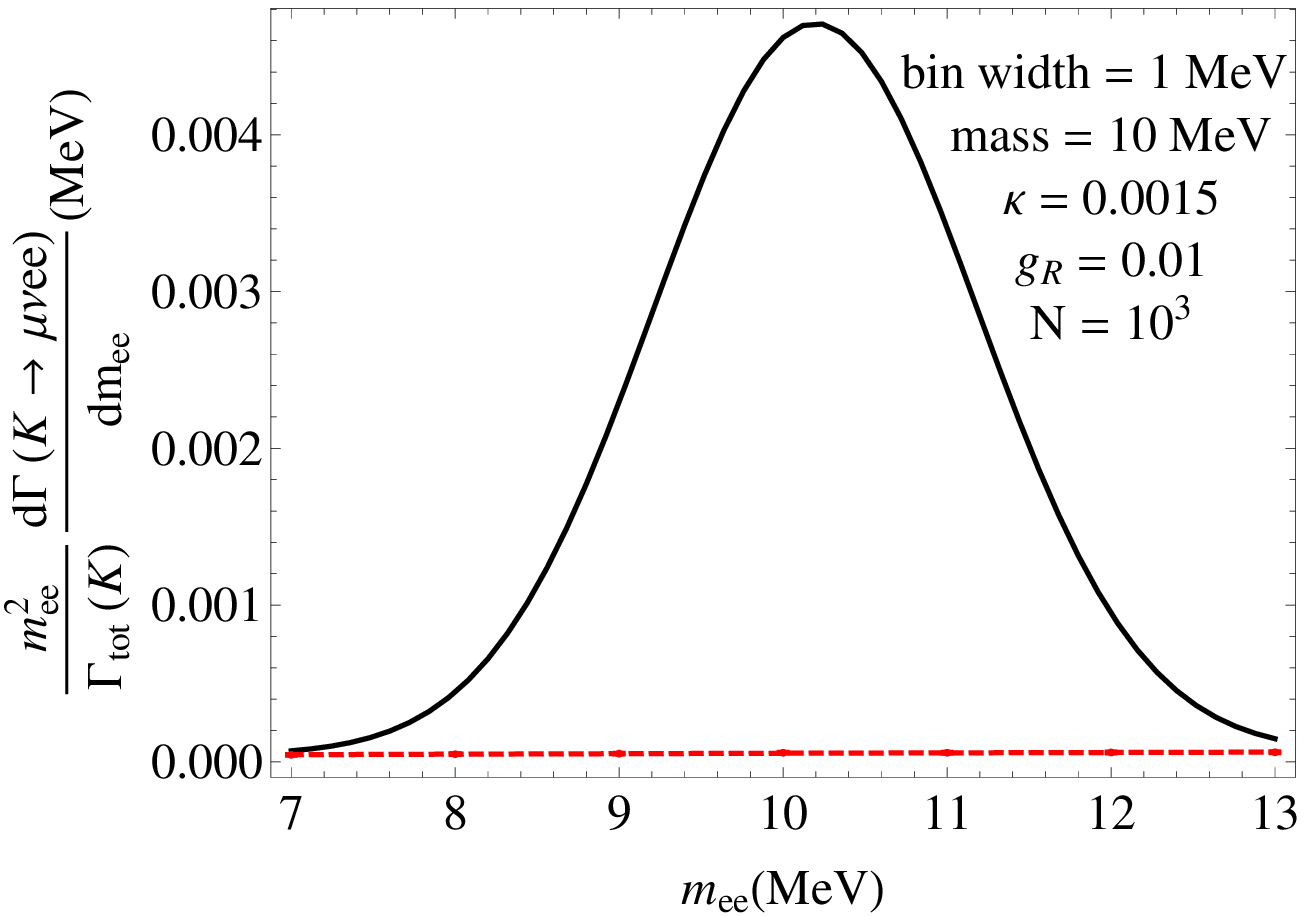}\hfil
\includegraphics[width = 70mm]{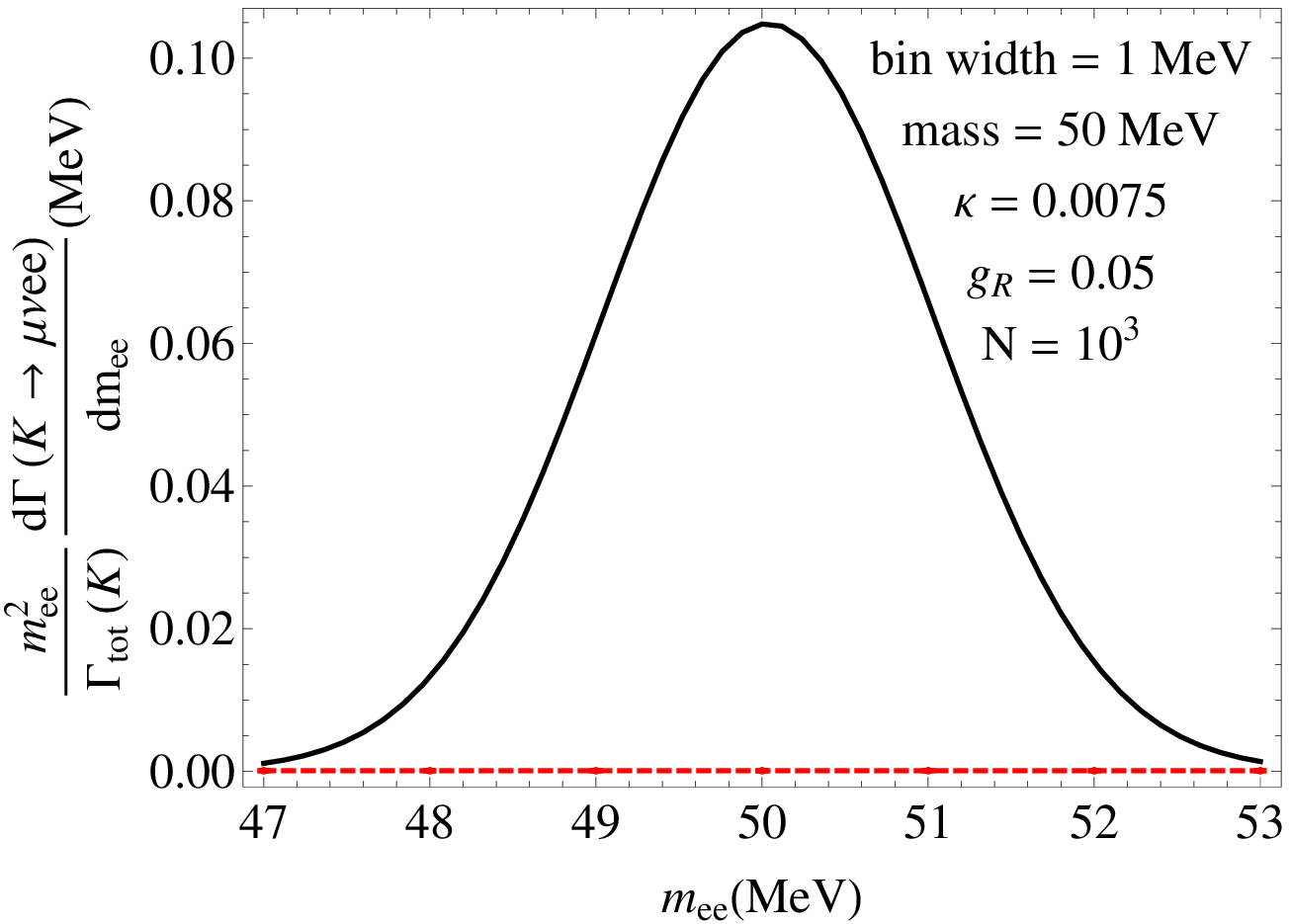}\hfil
\includegraphics[width = 70mm]{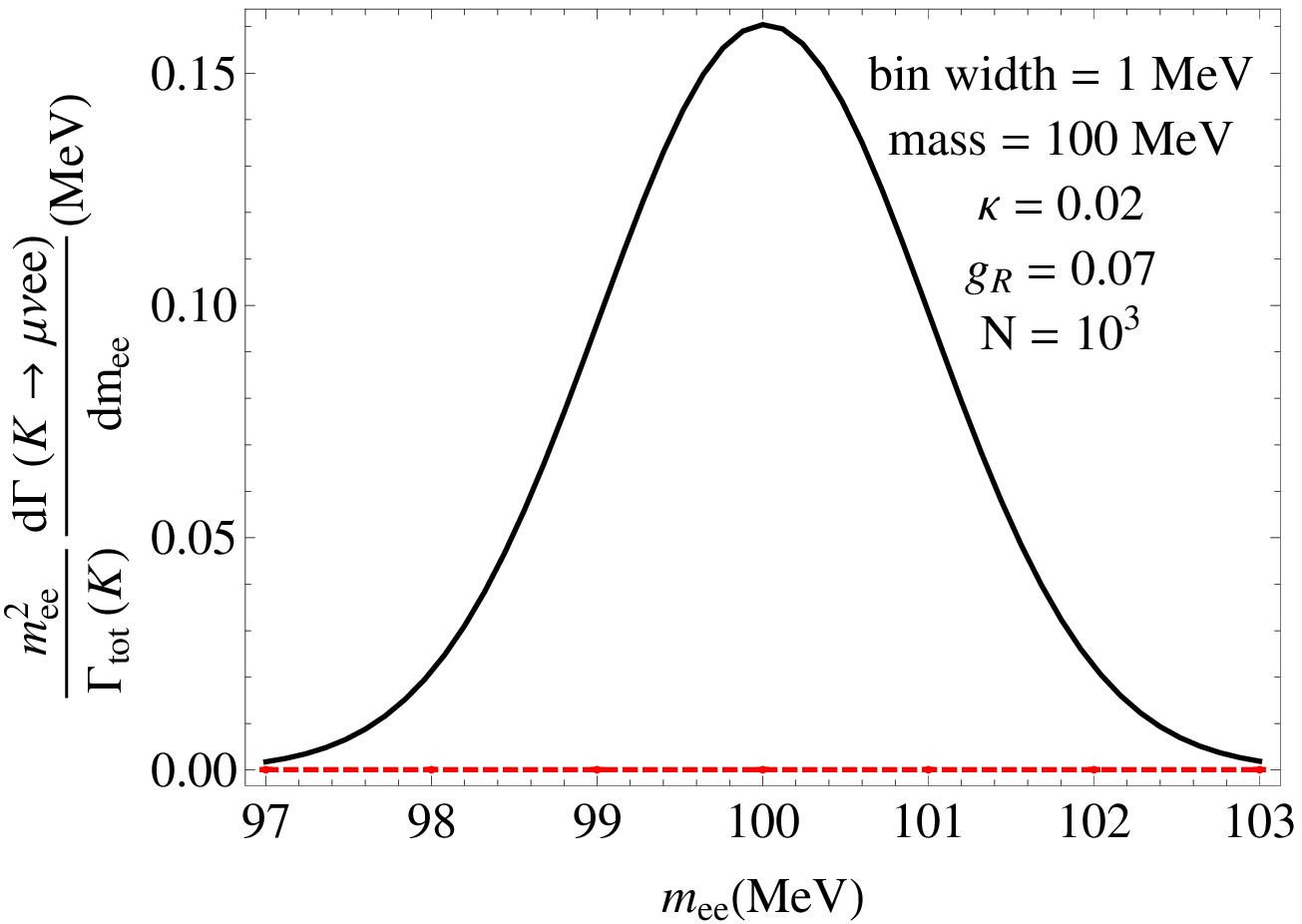}

\caption{QED prediction for $K^+\rightarrow \mu^+ +\nu_\mu +e^+ +e^-$ (red, dashed curve) and the prediction with the additional lepton-universality violating particle of Batell \textit{et al.}~(black curve).  Simulated data points with error bars accounting for the statistical uncertainty, anticipating $N=1000$ events per bin, are smaller than the width of the red line; $\kappa$ and $g_R$ are couplings in the Batell \textit{et al.}~model~\cite{Batell:2011qq}.}
\label{Batell}
\end{center}
\end{figure*}

JPARC could be quite sensitive to the lepton universality-violating particles present in proton charge radius puzzle explanations.  The model of Batell \textit{et al.}~\cite{Batell:2011qq} contains dark photon-esque particles that possess an additional, direct coupling to right-handed muons.  To account for this particle's effect on the amplitude $K^+\rightarrow \mu^+ +\nu_\mu+e^++e^-$, one modifies the electromagnetic coupling to particles other than the muon by $e\to \kappa e$, and modifies the photon propagator and muon coupling to 
\begin{align}
\frac{-i}{q^2} &\rightarrow \frac{-i}{q^2-m_{A'}^2+im_{A'}\Gamma},
\\
-ie\gamma^\mu &\rightarrow -i\kappa e\gamma^\mu-i\frac{g_R}{2}\gamma^\mu(1+\gamma^5),
\end{align}
where  $g_R$ is its additional right-handed coupling to muons.  Batell \textit{et al.}~give three examples for values of $m_{A'}$, $\kappa$, and $g_R$ that satisfy all present constraints.   The overall branching ratios for $K\to \mu \nu A' \to \mu \nu e^+ e^-$ range from a few tenths to about $10^{-4}$ for parameter values that they give.  Fig.~\ref{Batell} displays the predictions for their parameter values.  The figure shows the $e^+ e^-$ signal and the QED background for an experimental resolution of $1$ MeV, plotted \textit{vs.}~the $e^+ e^-$ mass.  This signal is several orders of magnitude greater than the one due to ``standard" dark photons and considerably larger than the QED background in the mass range of the new particle.  As can be seen by the expected experimental error bars, a dark photon with the Batell \textit{et al.} enhanced muon coupling should be detectable at JPARC.


\section{Our model contribution to $K^+ \rightarrow \mu^+ +\nu + e^- + e^+$.}			\label{sec:JPARC_us}


In our model~\cite{Carlson:2012pc} we considered new particles with polar and axial vector couplings to muons and protons.  The new particle's propagator is the same as in the Batell \textit{et al.}~model.  Our original model omitted a coupling to electrons and would not contribute to $K^+ \rightarrow \mu^+ +\nu + e^- + e^+$.  We now modify our model to include small couplings to electrons, being careful that the coupling strength respects the constraint placed by the uncertainty of the electron's anamolous magnetic moment.  According to the PDG's value of $\delta(g-2)_e$~\cite{Beringer:1900zz} and the latest $\delta(g-2)_\mu$ calculation~\cite{Aoyama:2012wk}, the new particle's coupling to the electron must be
\begin{align}
\label{momentratio}
\epsilon^2=\frac{\delta(g-2)_e}{\delta(g-2)_\mu}=1.1\times 10^{-4}
\end{align}
smaller than that of the muon.  

It is important to note that as long as the electron coupling is large enough that the new particle decays in the detector and is small enough that the decay width is narrower than the experimental resolution, the size of the bump than may be seen in the $e^+ e^-$ spectrum is independent of its value.  The allowed range of couplings is rather wide.

\begin{figure}[htbp]
\begin{center}

\includegraphics[width = 70mm]{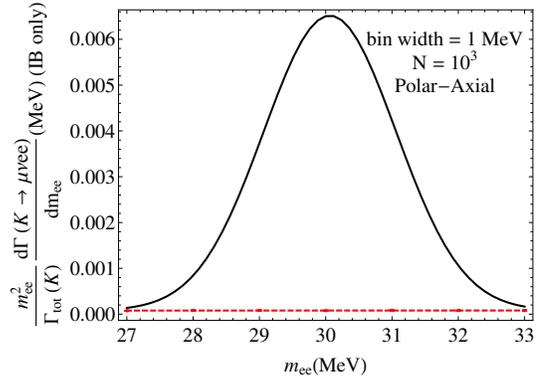}   

\caption{QED prediction for $K^+\rightarrow \mu^+ +\nu_\mu +e^+ +e^-$ (red, dashed curve) and the prediction with the additional 30 MeV lepton-universality violating polar/axial vector particle of our model~\cite{Carlson:2012pc} (black curve).  Simulated data points with error bars accounting for the statistical uncertainty are smaller than the width of the red line.}
\label{va}
\end{center}
\end{figure}

To account for this particle's contribution to the amplitude $K^+\rightarrow \mu^+ +\nu_\mu+e^++e^-$, the photon's propagator and charged fermion couplings are modified to
\begin{align}
\frac{-i}{q^2} &\rightarrow \frac{-i}{q^2-m_{A'}^2+im_{A'}\Gamma},
\\
-ie\gamma^\mu &\rightarrow -i\gamma^\mu\epsilon(C_V(m_{A'})+C_A(m_{A'})\gamma^5),
\end{align}
where $C_V(m_{A'})$ and $C_A(m_{A'})$ are the polar and axial vector couplings to muons calculated in~\cite{Carlson:2012pc}.  The $C_{V,A}$ are chosen, as described in~\cite{Carlson:2012pc}, so that the muonic Lamb shift is lowered by the correct amount, and so that the $(g-2)_\mu$ discrepancy is at its known value.  The values depend on the mass $m_{A\prime}$.  The parameter $\epsilon=1$ for the muon and is the square root of Eq.~(\ref{momentratio}) for the electron.  We assume the electron's value of $\epsilon$ can also be applied to the kaon.  See Fig.~\ref{va} for the prediction of a 30 MeV particle.  As mentioned in Sec.~\ref{sec:JPARC_QED}, $H^{\rho\nu}$ is very small and we only consider the IB contributions.  The peak is so large that omitting $H^{\rho\nu}$ will not affect the conclusion that this new particle will be detectable at JPARC.  The overall $K\to \mu \nu A' \to \mu \nu e^+ e^-$ branching ratio is few times $10^{-5}$ for polar/axial vector $A'$ and this mass.


\begin{figure}[htbp]
\begin{center}

\includegraphics[width = 70mm]{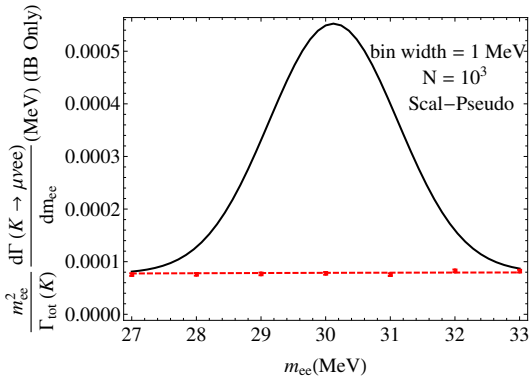}

\caption{QED prediction for $K^+\rightarrow \mu^+ +\nu_\mu +e^+ +e^-$ (red, dashed curve) and the prediction with the additional 30 MeV lepton-universality violating scalar/pseudoscalar particle of our model (black curve).  Data points are simulated and possess error bars accounting for the statistical uncertainty.}
\label{sp}
\end{center}
\end{figure}

The new particle may instead have scalar and pseudoscalar couplings to the muon and proton, and we model this also.  Its contribution to the amplitude $K^+\rightarrow \mu^+ +\nu_\mu+e^++e^-$ is found by modifying the photon's propagator as before and modifying the charged fermion couplings as:
\begin{align}
-ie\gamma^\mu\rightarrow -i\epsilon(C_S(m_{A'})+i C_P(m_{A'})\gamma^5).
\end{align}
Here, $C_S(m_{A'})$ and $C_P(m_{A'})$ are the scalar and pseudoscalar couplings to muons calculated in~\cite{Carlson:2012pc}.  For muons, $\epsilon=1$ and is smaller for other fermions.  See Fig.~\ref{sp} for the prediction of a 30 MeV particle with scalar and pseudoscalar couplings.  Here, we also only consider the IB contribution since $H^{\rho\nu}$ is so small.  The overall $K\to \mu \nu A' \to \mu \nu e^+ e^-$ branching ratio is few times $10^{-6}$ for this spin and mass.


\section{Conclusions}			\label{sec:JPARC}


Analyses of the proton radius conundrum have provoked suggestion of muon-electron non-universality that may also explain the $(g-2)_\mu$ puzzle but whose mass and couplings can otherwise be arranged to not contradict known constraints (see~\cite{TuckerSmith:2010ra,Batell:2011qq,Carlson:2012pc,Barger:2010aj,Barger:2011mt}).  Further tests should be made.  In particular, low mass new particles that couple to muons must show themselves throughout corrections to particle decays involving muons.

We have here focussed on what might be called leptonic Dalitz decays of the charged kaon, $K \to \mu \nu e^+ e^-$.  There is a calculated~\cite{Bijnens:1992en} and partly measured (in the higher $m_{ee}$ mass region) QED background~\cite{Poblaguev:2002ug}.  New light massive particles that couple well to muons, and to electrons at least enough to decay before leaving the detector apparatus, should stand out as bumps in the $e^+ e^-$ spectrum at the mass of the new particles.

An experiment like E36 at JPARC that anticipates $10^{10}$ kaon decays will see, by calculation, about $200,000$ QED produced $K \to \mu \nu e^+ e^-$ decays with a smooth and predictable background.  With these statistics and the masses and couplings of the new particles tuned to explain the proton radius discrepancy and to accommodate the $(g-2)_\mu$ discrepancy, the bumps in the $e^+ e^-$ spectrum will be very striking and very observable if the particles exist.


\begin{acknowledgments}

We thank Michael Kohl and Marc Vanderhaeghen for useful conversations.  We thank the National Science Foundation for support under Grant PHY-1205905.  B.~R. also thanks the DFG in the SFB 1044 for support.

\end{acknowledgments}

\bibliography{kaondecay}

\begin{thebibliography}{18}
\expandafter\ifx\csname natexlab\endcsname\relax\def\natexlab#1{#1}\fi
\expandafter\ifx\csname bibnamefont\endcsname\relax
  \def\bibnamefont#1{#1}\fi
\expandafter\ifx\csname bibfnamefont\endcsname\relax
  \def\bibfnamefont#1{#1}\fi
\expandafter\ifx\csname citenamefont\endcsname\relax
  \def\citenamefont#1{#1}\fi
\expandafter\ifx\csname url\endcsname\relax
  \def\url#1{\texttt{#1}}\fi
\expandafter\ifx\csname urlprefix\endcsname\relax\def\urlprefix{URL }\fi
\providecommand{\bibinfo}[2]{#2}
\providecommand{\eprint}[2][]{\url{#2}}

\bibitem[{\citenamefont{Pohl et~al.}(2010)\citenamefont{Pohl, Antognini, Nez,
  Amaro, Biraben et~al.}}]{Pohl:2010zza}
\bibinfo{author}{\bibfnamefont{R.}~\bibnamefont{Pohl}},
  \bibinfo{author}{\bibfnamefont{A.}~\bibnamefont{Antognini}},
  \bibinfo{author}{\bibfnamefont{F.}~\bibnamefont{Nez}},
  \bibinfo{author}{\bibfnamefont{F.~D.} \bibnamefont{Amaro}},
  \bibinfo{author}{\bibfnamefont{F.}~\bibnamefont{Biraben}},
  \bibnamefont{et~al.}, \bibinfo{journal}{Nature}
  \textbf{\bibinfo{volume}{466}}, \bibinfo{pages}{213} (\bibinfo{year}{2010}).

\bibitem[{\citenamefont{Antognini et~al.}(2013)\citenamefont{Antognini, Nez,
  Schuhmann, Amaro, Biraben et~al.}}]{Antognini:1900ns}
\bibinfo{author}{\bibfnamefont{A.}~\bibnamefont{Antognini}},
  \bibinfo{author}{\bibfnamefont{F.}~\bibnamefont{Nez}},
  \bibinfo{author}{\bibfnamefont{K.}~\bibnamefont{Schuhmann}},
  \bibinfo{author}{\bibfnamefont{F.~D.} \bibnamefont{Amaro}},
  \bibinfo{author}{\bibfnamefont{F.}~\bibnamefont{Biraben}},
  \bibnamefont{et~al.}, \bibinfo{journal}{Science}
  \textbf{\bibinfo{volume}{339}}, \bibinfo{pages}{417} (\bibinfo{year}{2013}).

\bibitem[{\citenamefont{Mohr et~al.}(2012)\citenamefont{Mohr, Taylor, and
  Newell}}]{Mohr:2012tt}
\bibinfo{author}{\bibfnamefont{P.~J.} \bibnamefont{Mohr}},
  \bibinfo{author}{\bibfnamefont{B.~N.} \bibnamefont{Taylor}},
  \bibnamefont{and} \bibinfo{author}{\bibfnamefont{D.~B.} \bibnamefont{Newell}}
  (\bibinfo{year}{2012}), \eprint{1203.5425}.

\bibitem[{\citenamefont{Jaeckel and Roy}(2010)}]{Jaeckel:2010xx}
\bibinfo{author}{\bibfnamefont{J.}~\bibnamefont{Jaeckel}} \bibnamefont{and}
  \bibinfo{author}{\bibfnamefont{S.}~\bibnamefont{Roy}},
  \bibinfo{journal}{Phys.Rev.} \textbf{\bibinfo{volume}{D82}},
  \bibinfo{pages}{125020} (\bibinfo{year}{2010}), \eprint{1008.3536}.

\bibitem[{\citenamefont{Tucker-Smith and Yavin}(2011)}]{TuckerSmith:2010ra}
\bibinfo{author}{\bibfnamefont{D.}~\bibnamefont{Tucker-Smith}}
  \bibnamefont{and} \bibinfo{author}{\bibfnamefont{I.}~\bibnamefont{Yavin}},
  \bibinfo{journal}{Phys.Rev.} \textbf{\bibinfo{volume}{D83}},
  \bibinfo{pages}{101702} (\bibinfo{year}{2011}), \eprint{1011.4922}.

\bibitem[{\citenamefont{Barger et~al.}(2011)\citenamefont{Barger, Chiang,
  Keung, and Marfatia}}]{Barger:2010aj}
\bibinfo{author}{\bibfnamefont{V.}~\bibnamefont{Barger}},
  \bibinfo{author}{\bibfnamefont{C.-W.} \bibnamefont{Chiang}},
  \bibinfo{author}{\bibfnamefont{W.-Y.} \bibnamefont{Keung}}, \bibnamefont{and}
  \bibinfo{author}{\bibfnamefont{D.}~\bibnamefont{Marfatia}},
  \bibinfo{journal}{Phys.Rev.Lett.} \textbf{\bibinfo{volume}{106}},
  \bibinfo{pages}{153001} (\bibinfo{year}{2011}), \eprint{1011.3519}.

\bibitem[{\citenamefont{Barger et~al.}(2012)\citenamefont{Barger, Chiang,
  Keung, and Marfatia}}]{Barger:2011mt}
\bibinfo{author}{\bibfnamefont{V.}~\bibnamefont{Barger}},
  \bibinfo{author}{\bibfnamefont{C.-W.} \bibnamefont{Chiang}},
  \bibinfo{author}{\bibfnamefont{W.-Y.} \bibnamefont{Keung}}, \bibnamefont{and}
  \bibinfo{author}{\bibfnamefont{D.}~\bibnamefont{Marfatia}},
  \bibinfo{journal}{Phys.Rev.Lett.} \textbf{\bibinfo{volume}{108}},
  \bibinfo{pages}{081802} (\bibinfo{year}{2012}), \bibinfo{note}{3 pages, 3
  figures. Version to appear in PRL}, \eprint{1109.6652}.

\bibitem[{\citenamefont{Batell et~al.}(2011)\citenamefont{Batell, McKeen, and
  Pospelov}}]{Batell:2011qq}
\bibinfo{author}{\bibfnamefont{B.}~\bibnamefont{Batell}},
  \bibinfo{author}{\bibfnamefont{D.}~\bibnamefont{McKeen}}, \bibnamefont{and}
  \bibinfo{author}{\bibfnamefont{M.}~\bibnamefont{Pospelov}},
  \bibinfo{journal}{Phys.Rev.Lett.} \textbf{\bibinfo{volume}{107}},
  \bibinfo{pages}{011803} (\bibinfo{year}{2011}), \eprint{1103.0721}.

\bibitem[{\citenamefont{Carlson and Rislow}(2012)}]{Carlson:2012pc}
\bibinfo{author}{\bibfnamefont{C.~E.} \bibnamefont{Carlson}} \bibnamefont{and}
  \bibinfo{author}{\bibfnamefont{B.~C.} \bibnamefont{Rislow}},
  \bibinfo{journal}{Phys. Rev.} \textbf{\bibinfo{volume}{D86}},
  \bibinfo{pages}{035013} (\bibinfo{year}{2012}), \eprint{1206.3587}.

\bibitem[{\citenamefont{Pang et~al.}(1973)\citenamefont{Pang, Hildebrand,
  Cable, and Stiening}}]{Pang:1989ut}
\bibinfo{author}{\bibfnamefont{C.}~\bibnamefont{Pang}},
  \bibinfo{author}{\bibfnamefont{R.}~\bibnamefont{Hildebrand}},
  \bibinfo{author}{\bibfnamefont{G.}~\bibnamefont{Cable}}, \bibnamefont{and}
  \bibinfo{author}{\bibfnamefont{R.}~\bibnamefont{Stiening}},
  \bibinfo{journal}{Phys.Rev.} \textbf{\bibinfo{volume}{D8}},
  \bibinfo{pages}{1989} (\bibinfo{year}{1973}).

\bibitem[{\citenamefont{Djalali}(2012)}]{Djalali:2012zz}
\bibinfo{author}{\bibfnamefont{C.}~\bibnamefont{Djalali}}
  (\bibinfo{collaboration}{TREK Collaboration}), \bibinfo{journal}{AIP
  Conf.Proc.} \textbf{\bibinfo{volume}{1423}}, \bibinfo{pages}{297}
  (\bibinfo{year}{2012}), \bibinfo{note}{and Michael Kohl, private
  communication.}

\bibitem[{\citenamefont{Kohl}(2013)}]{Jparc}
\bibinfo{author}{\bibfnamefont{M.}~\bibnamefont{Kohl}} (\bibinfo{year}{2013}),
  \bibinfo{note}{private communication}.

\bibitem[{\citenamefont{Beringer et~al.}(2012)}]{Beringer:1900zz}
\bibinfo{author}{\bibfnamefont{J.}~\bibnamefont{Beringer}} \bibnamefont{et~al.}
  (\bibinfo{collaboration}{Particle Data Group}), \bibinfo{journal}{Phys.Rev.}
  \textbf{\bibinfo{volume}{D86}}, \bibinfo{pages}{010001}
  (\bibinfo{year}{2012}).

\bibitem[{\citenamefont{Bijnens et~al.}(1993)\citenamefont{Bijnens, Ecker, and
  Gasser}}]{Bijnens:1992en}
\bibinfo{author}{\bibfnamefont{J.}~\bibnamefont{Bijnens}},
  \bibinfo{author}{\bibfnamefont{G.}~\bibnamefont{Ecker}}, \bibnamefont{and}
  \bibinfo{author}{\bibfnamefont{J.}~\bibnamefont{Gasser}},
  \bibinfo{journal}{Nucl. Phys.} \textbf{\bibinfo{volume}{B396}},
  \bibinfo{pages}{81} (\bibinfo{year}{1993}), \eprint{hep-ph/9209261}.

\bibitem[{\citenamefont{Beranek et~al.}(2013)\citenamefont{Beranek, Merkel, and
  Vanderhaeghen}}]{Beranek:2013yqa}
\bibinfo{author}{\bibfnamefont{T.}~\bibnamefont{Beranek}},
  \bibinfo{author}{\bibfnamefont{H.}~\bibnamefont{Merkel}}, \bibnamefont{and}
  \bibinfo{author}{\bibfnamefont{M.}~\bibnamefont{Vanderhaeghen}}
  (\bibinfo{year}{2013}), \eprint{1303.2540}.

\bibitem[{\citenamefont{Krishna and Mani}(1972)}]{Krishna:1972za}
\bibinfo{author}{\bibfnamefont{S.}~\bibnamefont{Krishna}} \bibnamefont{and}
  \bibinfo{author}{\bibfnamefont{H.}~\bibnamefont{Mani}},
  \bibinfo{journal}{Phys.Rev.} \textbf{\bibinfo{volume}{D5}},
  \bibinfo{pages}{678} (\bibinfo{year}{1972}).

\bibitem[{\citenamefont{Poblaguev et~al.}(2002)\citenamefont{Poblaguev, Appel,
  Atoyan, Bassalleck, Bergman et~al.}}]{Poblaguev:2002ug}
\bibinfo{author}{\bibfnamefont{A.}~\bibnamefont{Poblaguev}},
  \bibinfo{author}{\bibfnamefont{R.}~\bibnamefont{Appel}},
  \bibinfo{author}{\bibfnamefont{G.}~\bibnamefont{Atoyan}},
  \bibinfo{author}{\bibfnamefont{B.}~\bibnamefont{Bassalleck}},
  \bibinfo{author}{\bibfnamefont{D.}~\bibnamefont{Bergman}},
  \bibnamefont{et~al.}, \bibinfo{journal}{Phys. Rev. Lett.}
  \textbf{\bibinfo{volume}{89}}, \bibinfo{pages}{061803}
  (\bibinfo{year}{2002}), \eprint{hep-ex/0204006}.

\bibitem[{\citenamefont{Aoyama et~al.}(2012)\citenamefont{Aoyama, Hayakawa,
  Kinoshita, and Nio}}]{Aoyama:2012wk}
\bibinfo{author}{\bibfnamefont{T.}~\bibnamefont{Aoyama}},
  \bibinfo{author}{\bibfnamefont{M.}~\bibnamefont{Hayakawa}},
  \bibinfo{author}{\bibfnamefont{T.}~\bibnamefont{Kinoshita}},
  \bibnamefont{and} \bibinfo{author}{\bibfnamefont{M.}~\bibnamefont{Nio}},
  \bibinfo{journal}{Phys.Rev.Lett.} \textbf{\bibinfo{volume}{109}},
  \bibinfo{pages}{111808} (\bibinfo{year}{2012}), \eprint{1205.5370}.

\end{thebibliography}

\end{document}